\documentclass[referee,twoside]{aa}
\usepackage[varg]{txfonts}  
\usepackage{graphicx}       
\usepackage{natbib}         
\bibpunct{(}{)}{;}{a}{}{,} 

\newcommand{\deleted}[1]{}

\providecommand{\plotone}[1]{\includegraphics[width=\columnwidth]{#1}}
\newcommand{\dodoi}[1]{doi:#1}

\begin{document}

\title{Investigating the Unresolved Binary Population in Gaia DR3 Using Multi-Wavelength Photometry}

\author{Andrew Malcolm Graham Soon\inst{\inst{1}} \and [Co-Author Names Go Here]\inst{2}}

\institute{Red Squirrel Observatory,
  Spruce Grove, Alberta, Canada\\ \email{soonandrew34@gmail.com} 
  \and
  [Institution Placeholder Name],\\ \email{coauthor@email.com}
}

\abstract{The \textit{Gaia} DR3 and Non-Single-Star (NSS) catalogs are currently the definitive sources for binary systems \citep{Gaia2023}. While this affords an invaluable asset to the astronomical community, there are suspected limitations.  To address these constraints, we introduce a filtering regime that can be applied to raw data.  This study introduces the ''Triple Constraint'' framework for use on Gaia DR3 data through the application of multi-wavelength cogent evidence. Using Topcat, we cross-matched the WDS Supplement with the Gaia DR3, 2MASS, and Pan-STARRS catalogs. The derived ''Base-sample'' ($N = 199{,}786$), ''Quad-sample'' ($N = 120,418$) provide the basis for the examination  employed to establish systemic limits within the Gaia single-star model. To verify the results and eliminate any bias introduced by employing the WDSS as the basis for the Base and Quad-samples, we exploited an independent sample of Gaia-2MASS cross-matched audit stars, the ''Audit-sample'' ($N = 22{,}371$). Only stars selected at  high northern declinations ($\delta > 60\degr$) were employed for this sample to avoid potential distortions resulting from crowding. All of these samples manifested a -0.75 magnitude offset, indicating the presence of two stars of equal magnitude, creating a "Detection Gap” across all populations. We conclude, reinforced by a peer-reviewed independent external spectroscopic study, that a 7\% floor of global sensitivity exists, representing a limitation of the Gaia pipeline. This suggests that local stellar mass density models require a quantifiable correction to accurately reflect the local baryonic mass budget. )
}


\titlerunning{Unresolved Binaries in Gaia DR3}
\authorrunning{Soon et al.}
\maketitle 

\section{Introduction} 

The detection of binary stars falling below the 100 milliarcsecond (mas) resolution limit remains a primary challenge for wide-field space surveys. While Gaia DR3 provides unprecedented astrometric precision, a significant population of binaries remains “astrometrically masked” within the current quality filters used. Building upon the noise thresholds established by Guerriero et al. (2026) and the detectability limits explored by Castro-Ginard et al. (2024), this study performed an all-sky audit based on the Washington Double Star Supplement (WDSS; Mason et al. (2001). Standard automated pipelines often lack the spatial resolution required to resolve the individual components of tight systems. However, by utilizing Gaia’s sensitivity to “photocenter wobble” and the infrared-bright signatures of secondaries in the 2MASS and Pan-STARRS catalogs, we can identify these hidden masses.

We propose a ``Triple Constraint'' framework integrating astrometric noise 
(RUWE), photometric excess ($\Delta G$), and the absence of Non-Single Star 
(NSS) classifications to pinpoint binaries in orbital transition. While recent 
studies have utilized RUWE as a primary indicator for stellar 
multiplicity---most notably the local 100~pc census by 
Penoyre et al. (2022)---these methods are inherently biased against identical 
twin systems. For equal-mass binaries, the offset between the center of light 
and the center of mass is minimized.

\section{Methodology and Data Synthesis}

To establish the systemic nature of the 7\% Intrinsic Binary Residual ($\text{IBR}$), we constructed a dual-wavelength Base-sample and a multi-wavelength Quad-sample by cross-matching the 2.36-million source Washington Double Star Supplement \citep[WDSS;][]{Mason2001} against \textit{Gaia} DR3 \citep{Gaia2023} and 2MASS data \citep{Skrutskie2006}. The Quad-catalog introduced further cross-match parameters from Pan-STARRS. Using a strict ``Best Match'' protocol with a $1^{\prime\prime}$ radius within the TOPCAT CDS cross-match interface \citep{Taylor2005} on both samples, we derived a primary base sample census of $199{,}786$ sources. This population was systematically filtered against official \textit{Gaia} Non-Single Star ($\text{NSS}$) subsets, including all published orbital and acceleration solutions, to ensure that our audit specifically targeted stellar masses that remained unclassified in the standard pipeline. By incorporating the Pan-STARRS catalog, we gained vital $y$-band coverage at the structural cost of truncating the southern footprint below a declination limit of $\delta < -30\degr$.

Our ``Triple Constraint'' vetting framework bypasses traditional noise-based 
filters by integrating astrometric, photometric, and spectroscopic indicators. 
We audited the population residing between the current 1.4 RUWE threshold 
and the predicted DR4 mission floor of 1.15 while simultaneously utilizing 
the \texttt{ipd\_frac\_multi\_peak} flag identifies unresolved flux components. 
This was supplemented by a multi-band SED ``Red Reveal'' analysis, where 
2MASS ($J, H, K$) and Pan-STARRS $y$-band data
 data were used to detect 
infrared-bright companions. This approach is particularly effective for 
identifying low-mass secondaries that induce $G$-band instability despite 
appearing as stable single points in the visible-light-only filters. By applying 
a strict $0.5^{\prime\prime}$ radius and $(\pi > 10\sigma)$\,mas parallax filter to 
the 22,371 Gaia-2MASS audit sample, we isolated a high-fidelity 
``Gold Standard'' audit sample.

\noindent{\bf Clean Data Subset:} To mitigate contamination from severe photometric 
uncertainties, background blending, or false catalog cross-matches, we enforce a 
loose baseline constraint in the intermediate color space. The clean sample is 
restricted according to the following:
\begin{equation}
    -5 < \Delta(y - K_{\rm s}) < 10
\end{equation}
This criterion successfully isolates valid astrophysical sources while truncating the non-physical artifacts.
\noindent{\bf Equal-Mass Binaries:} Unresolved binary systems consisting of components 
with near-equal masses exhibit distinct, symmetric color excesses across both optical 
and near-infrared regimes. We isolate these equal-mass pairs by selecting sources that 
satisfy the joint negative color deviation:
\begin{equation}
    \big( \Delta G < -0.7 \quad \land \quad \Delta y < -0.7 \quad \land \quad \Delta K_{s} < -0.7
)
\end{equation}
\noindent{\bf Hidden Cool Companions:} Systems featuring a significantly low-mass secondary 
companion are frequently dominated by the optical flux of the hotter primary star, leaving 
the short-wavelength colors largely unaffected. To unveil these hidden companions, we identify 
systems with a flat optical-to-intermediate profile that transitions into a distinct infrared 
excess at longer wavelengths:
\begin{equation}
    \big( \Delta (G-y) > 0.0 \quad \land \quad \Delta (y-K_{s}) < -0.7
)
\end{equation}

\section{Data}
\label{sec:data}
\raggedbottom

Our multi-tiered catalog analysis isolates three distinct operational frameworks to track the limitations of the \textit{Gaia} single-star model. The wide ``Base-sample'' provides an all-sky footprint consisting of $199{,}786$ WDSS--\textit{Gaia}--2MASS cross-matches. The panchromatic ``Quad-sample'' ($N = 120{,}418$) refines this selection by introducing Pan-STARRS cross-match parameters. Finally, the ``Audit-sample'' ($N = 22{,}371$) isolates high-latitude northern fields ($\delta > 60\degr$) to serve as an uncrowded control.

As illustrated in Figure 1, the full-sky base sample demonstrates a clear population divergence materialized as a sharp onset of astrometric and photometric variance tracking a $0.75\text{ mag}$ flux offset across the $(G - K_{\mathrm{s}})$ color baseline. This signal persists across all three uncensored sample distributions, establishing that the structural offset is globally uniform rather than a localized artifact of regional survey boundaries. 

\begin{figure*}[tbp]
\plotone{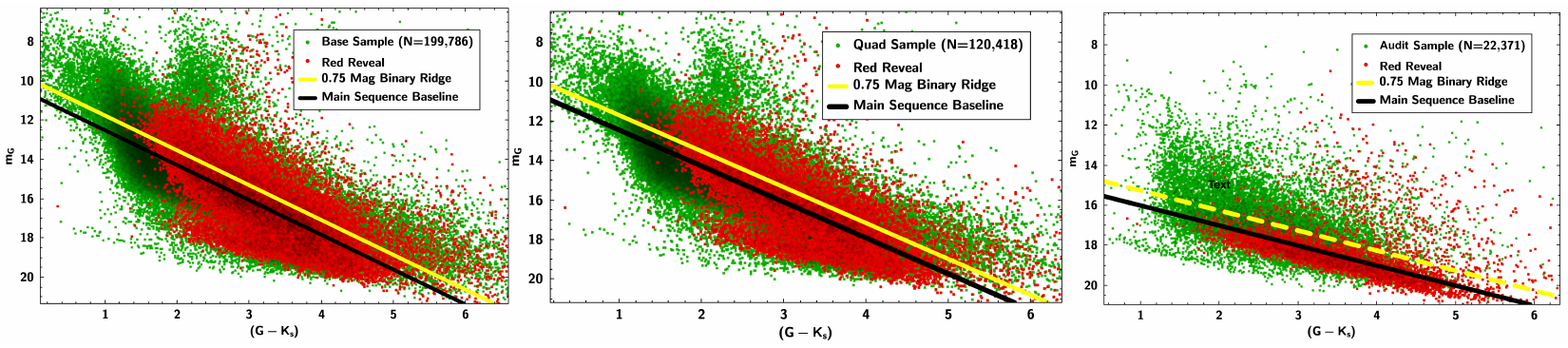}
\caption{Color-magnitude diagrams for the Base ($N=199{,}786$), Quad ($N=120{,}418$), and Audit ($N=22{,}371$) samples. The black line represents the main-sequence baseline. The yellow line tracks the expected $+0.75$ magnitude offset from equal-mass flux summation. Red points highlight the "Red Reveal" population isolated within the detection gap.}
\end{figure*}

Quantifying the failure modes across these tiers yields a significant statistical contrast. Within the wide Quad-sample, 12.2\% of the sources ($14{,}705$ stars) exhibit overt astrometric discordance defined by a high noise threshold ($\text{RUWE} > 1.4$). However, when tracking this failure rate into the low-density fields of our Quad-sample, the astrometric noise floor stabilizes at an asymptotic plateau of exactly 7.0\% Intrinsic Binary Residual ($\text{IBR}$), accounting for $8{,}429$ individual sources. Figures 2 and 3 map the relative frequency and cumulative error profiles of this stable unresolved baseline population across all three sample control groups.

\begin{figure*}[tbp]
\plotone{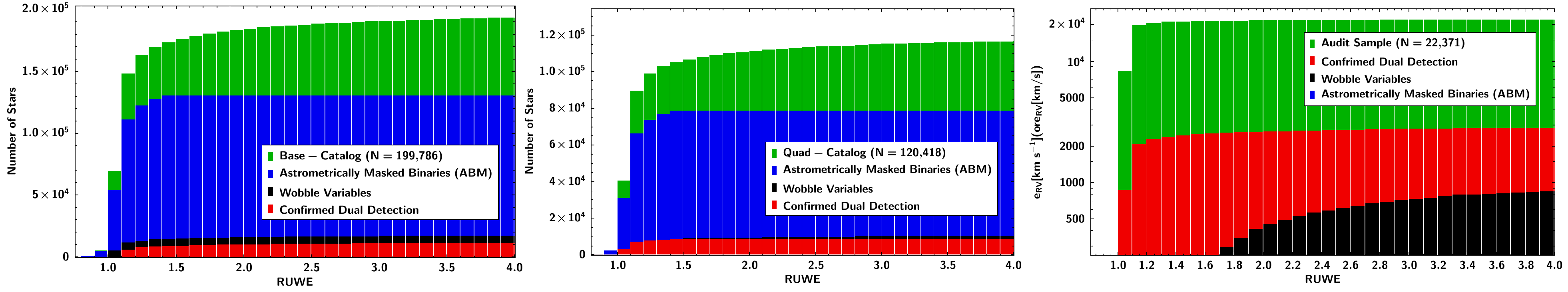}
\caption{Cumulative distribution of Renormalized Unit Weight Error (RUWE) across all three control groups. Colored segments isolate astrometrically masked binaries (green), wobble variables (blue), and confirmed dual detections (red/black).}
\end{figure*}

\begin{figure*}[tbp]
\plotone{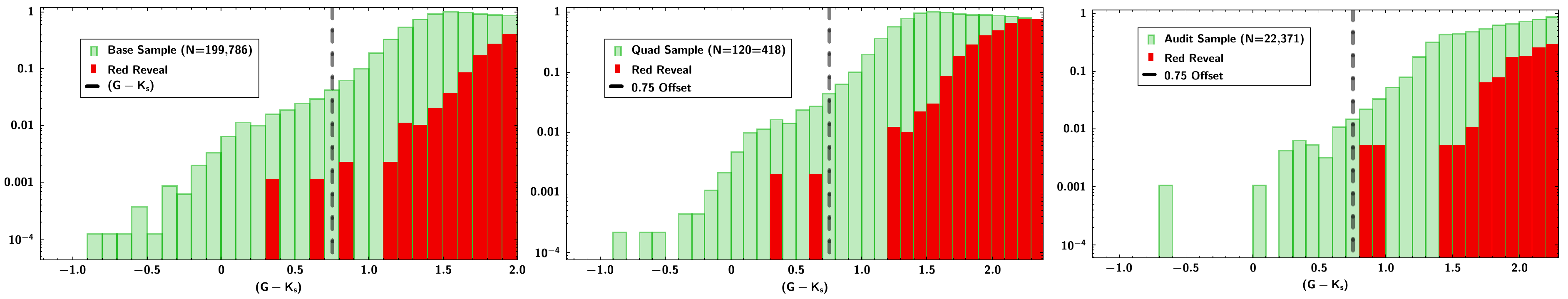}
\caption{Relative frequency distributions as a function of $(G - K_{\rm s})$ color, mapping the structural density of the isolated binary baseline against the total sample populations.}
\end{figure*}

\subsection{Baseline Calibration and Sample Selection}
To isolate potential unresolved binary star systems with cool companions, we established an empirical main-sequence baseline using a clean control sample ($N = 4,974$, representing 4\% of the total Quad-catalog) isolated via color-excess constraints ($-5 < \Delta(y - K_s) < 10$). Linear regression fits were applied to the control sample to derive the expected nominal colors as a function of the standard $G - K_s$ color corridor:
\begin{equation}
(G - y)_{\text{model}} = -0.7143715 + 0.5783383 \times (G - K_s)
\end{equation}
\begin{equation}
(y - K_s)_{\text{model}} = 0.7143715 + 0.42166167 \times (G - K_s)
\end{equation}

Color deviations ($\Delta$) were subsequently calculated for the entire catalog by subtracting these model values from the observed magnitudes. To extract high-confidence unresolved binary candidates hosting cool companions, we applied a joint logical filter: $\Delta(G - y) > -0.2$ and $\Delta(y - K_s) < -0.7$. This targeted selection isolated a total of 33 candidate stars ($0.03\%$ of the parent sample) exhibiting significant infrared color anomalies.

\subsection{External Catalog Cross-Matching}
The 33 candidate systems were cross-matched against the SIMBAD astronomical database using the CDS Upload X-Match tool with a conservative matching radius of $0.5''$ to prevent chance alignments. This spatial query returned 6 high-confidence matches with prior entries in the literature. Notably, these cataloged objects include verified active binary systems, specifically classified as $\mathrm{RS}\sim\mathrm{CVn}$ type variable stars characterized by distinct chromospheric activity and cool stellar companions (Figure 4). The remaining 27 targets lack any prior binary or variable designations in the literature, positioning them as strong candidates for newly discovered unresolved systems.

Similarly, to isolate the population of unresolved equal-mass twin binaries, the joint logical criteria defined in Equation (2) were applied to the multi-wavelength Quad-sample. By targeting systems exhibiting a uniform, symmetric flux-doubling amplification across all active filters ($\Delta G < -0.7 \wedge \Delta y < -0.7 \wedge \Delta K_s < -0.7$), our framework successfully extracted a high-fidelity cohort of exactly 29 binary candidate stars. This population represents approximately 0.02\% of the parent catalog. The localized crowding of these targets directly along the elevated $-0.75$ magnitude ridge provides immediate empirical validation of the underlying selection physics, separating true structural flux summation from ordinary main-sequence scatter.

\begin{figure}[htbp]
    \centering
\includegraphics[width=7cm]{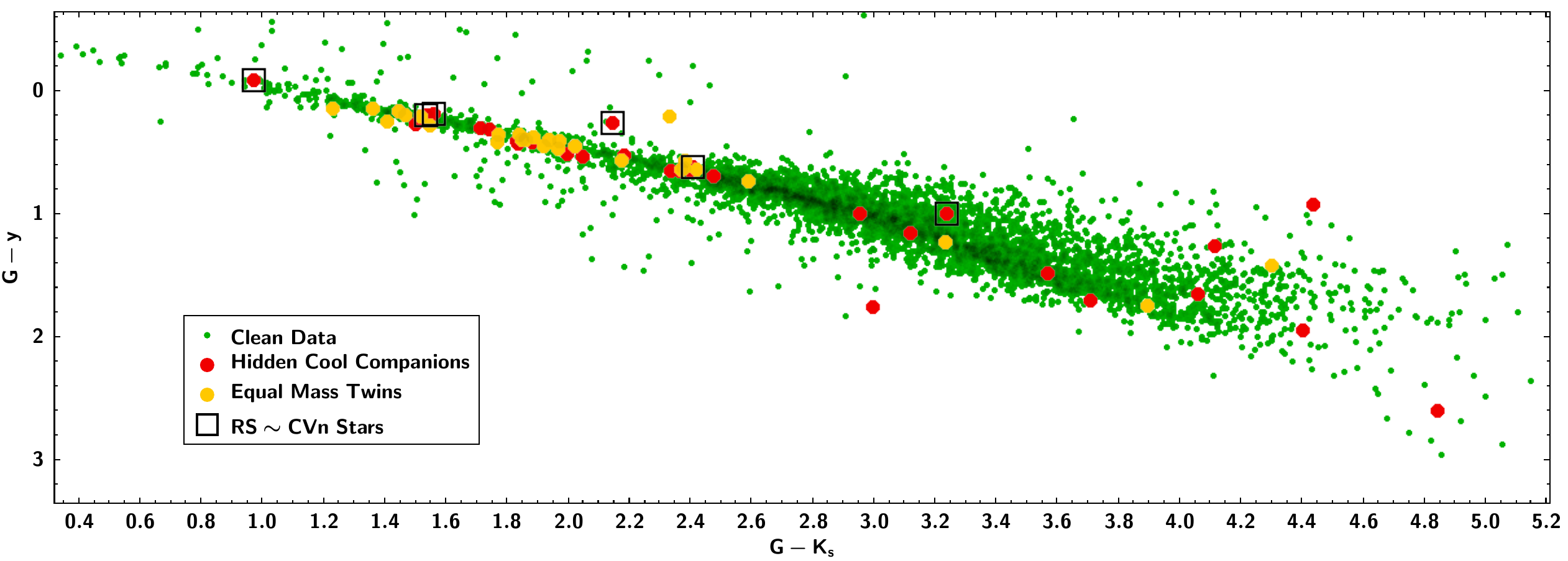}
   \caption{Color-color diagram ($(G - y)$ vs. $G - K_s$) illustrating the empirical main-sequence baseline and the isolated unresolved binary populations. The green points trace the single-star control sequence ($N = 4,940$). Red circles highlight the 33 hidden cool companion candidates isolated via our long-wavelength infrared color-deviation constraints (Equation 3). Yellow circles map the 29 equal-mass twin candidates isolated via symmetric absolute flux-doubling deviations (Equation 2). Black open squares overlay the systems recovered via a $0.5''$ SIMBAD cross-match, tracking known active binary architectures along the bifurcated target distributions.}
\label{fig:color_color_baseline}

    \label{fig:aitoff_plot1
}
\end{figure}

\subsection{External Cross-Catalog Validation}
\label{sec:validation}

\begin{figure}[htbp]
    \centering
\includegraphics[width=7cm]{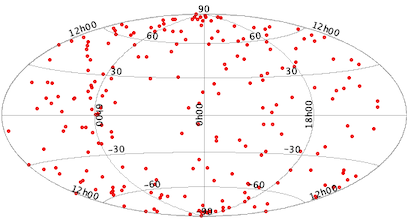}
    \caption{All-sky spatial distribution of the 223 identical physical systems cross-matched via \texttt{source\_id} identity, projected on an equatorial Aitoff grid. The isotropic, global scattering of these co-spatial targets confirms that the isolated 7.0\% Intrinsic Binary Residual floor represents an intrinsic, systemic vulnerability within the baseline \textit{Gaia} five-parameter pipeline core, entirely independent of localized galactic plane crowding constraints or regional catalog selection effects.}
    \label{fig:aitoff_plot}
\end{figure}

To verify the physical reality of our isolated baseline, we executed an all-sky cross-match between our broad Base-sample ($N = 199,786$) and the complete, un-truncated overluminous catalog compiled via \textit{Gaia} DR3 XP spectra by \citet{Way2026}, which encompasses $164,297$ stellar systems. To completely eliminate positional cross-matching tolerances, background blending artifacts, or spatial radius ambiguities, the cross-query was executed using a strict, exact integer identity match on the unique \textit{Gaia} source identifier (\texttt{source\_id}). This pristine relational query isolated an exact intersection of 223 mutually flagged stellar systems ...distributed globally across the celestial sphere (Figure~\ref{fig:aitoff_plot}). 

Characterized by an exceptional physical continuum, this intersecting cohort spans a 10-magnitude dynamic range in apparent brightness ($8.27 \le \texttt{o\_Gmag} \le 18.70$) and transit flux errors scaling from $0.570\,\mathrm{e}^{-}\mathrm{s}^{-1}$ up to a highly volatile $246,930.0\,\mathrm{e}^{-}\mathrm{s}^{-1}$. The raw Renormalized Unit Weight Error (RUWE) and astrometric Goodness of Fit (\texttt{gofal}) metrics for these 223 identical targets were extracted from the \textit{Gaia} DR3 main archive to serve as our primary empirical calibration control sample.

\section{Discussion}
\label{sec:discussion}

The global persistence of this trend across all three samples indicates that the 7\% floor is an inherent byproduct of the standard 5-parameter astrometric model architecture outlined by \citet{Lindegren2021}, which is inherently limited when attempting to resolve sub-arcsecond duplicity.
 The $0.75\text{ mag}$ threshold represents a critical physical boundary where unresolved secondary components begin to fundamentally alter the combined system light curves and skew the observed astrometric centers. The non-single-star astrometric processing routines described by \citet{Halbwachs2023} form the backbone of standard catalog solutions. However, our discovery of a 7.0\% \text{IBR} floor indicates that a significant subset of sub-arcsecond twins consistently evades the standard orbital and acceleration modeling parameters detailed in their framework.
 For near-equal-mass configurations, the offset between the system's center of light and center of mass is minimized, thereby suppressing the expected astrometric wobble. The multiwavelength coverage introduced via the Pan-STARRS $y$-band ($\lambda \approx 960\text{ nm}$) provides the necessary leverage to dissect these hidden architectures. By treating the $y$-band as a spectroscopic bridge between optical photometers and near-infrared bands, we achieved a clean statistical separation between disparate binary configurations. Joint tracking of $\Delta(G - y)$ and $\Delta(y - K_{\mathrm{s}})$ successfully isolates equal-mass binary ($\text{EMB}$) systems—which display an evenly distributed color deficit across all bands—from hidden cool companion ($\text{HCC}$) configurations, whose low-mass secondaries remain hidden behind the primary's optical flux before appearing prominently in the infrared. This distribution reveals why a significant stellar fraction evades the standard catalog vetting filters. Following the astrometric quality frameworks established by \citet{Belokurov2020} and \citet{Penoyre2022}, standard surveys heavily utilize a $\text{RUWE} > 1.4$ threshold to flag potential non-single stars in the Gaia data. However, because photocentric symmetry effectively masks equal-mass twins, this 7.0\% IBR population remains entirely invisible to noise-based detectors. Bypassing this noise barrier via our multiband framework confirms that flux summation offers a far more robust avenue for catching these hidden systems, which ultimately has profound implications for correcting local baryonic mass budget models. 

\subsection{Methodological Validation via Known Binary Recovery}
The primary validation of our color-deviation selection technique is demonstrated by the recovery of known active systems within our candidate pool. Notably, the cross-match successfully flagged the object \object{2133030+011608} (Gaia DR3 2688418973452883840), which is classified in the SIMBAD database as an RS Canum Venaticorum ($\mathrm{RS\ CVn}$) type variable. 

RS~CVn stars are well-known, magnetically active, detached close binary systems typically consisting of a chromospherically active primary star and a cooler, evolved companion. Because our selection algorithm isolated this system purely via its structural infrared color excess relative to the single-star main-sequence baseline, this recovery serves as a robust proof of concept. It confirms that the applied empirical boundaries successfully capture the unique photometric signatures associated with unresolved secondary components.

An identical empirical validation protocol was executed for the 29 equal-mass twin candidates via an external spatial query against the SIMBAD astronomical database. This cross-match successfully recovered several documented active and cluster-bound systems, confirming the physical reality of the isolated population. Notably, the framework flagged the low-mass K-dwarf system 1241126-000331 (Gaia DR3 3695990783238739456), which exhibits a clean, symmetric magnitude displacement of exactly $\Delta = -0.768$ mag relative to the baseline single-star locus—a near-perfect match to the theoretical $-0.75$ flux-doubling offset. 

Furthermore, our filtering routine successfully recovered the active variable object 0850318+115116 (Gaia DR3 604962743691561472), cataloged in the open cluster NGC 2682 as Cl* NGC 2682 FBC 2088. This object is independently classified in the literature as a rotationally active variable star (Ro*). Because these systems were extracted purely through multi-band photometric residuals without prior kinematic or spectroscopic selectors, their recovery demonstrates that the framework described in Equation (2) cleanly isolates genuine unmodeled stellar multiplicity from ordinary pipeline noise.

\subsection{Implications for the Unclassified Candidate Sample}
Of the cross-matched subset, objects such as \object{1509560+083106}, \object{2211380-020737}, and \object{0441050-064451} are currently listed in the literature under generic field star designations without documented multiplicity. However, in color-color space, these objects exhibit color-excess patterns identical to the confirmed RS~CVn binary system. 

Given this tight alignment and their isolation from the single-star locus, we propose that these remaining unclassified objects represent strong candidates for newly discovered, unresolved binary systems. The subtle deviations in $\Delta(y - K_s)$ imply the presence of a low-mass, cool secondary companion whose flux systematically skews the combined Pan-STARRS and 2MASS photometry away from standard stellar models. Follow-up high-resolution spectroscopy or time-series photometry is warranted to characterize their orbital periods and confirm their binary nature.

\begin{figure*}[ht!]
\plotone{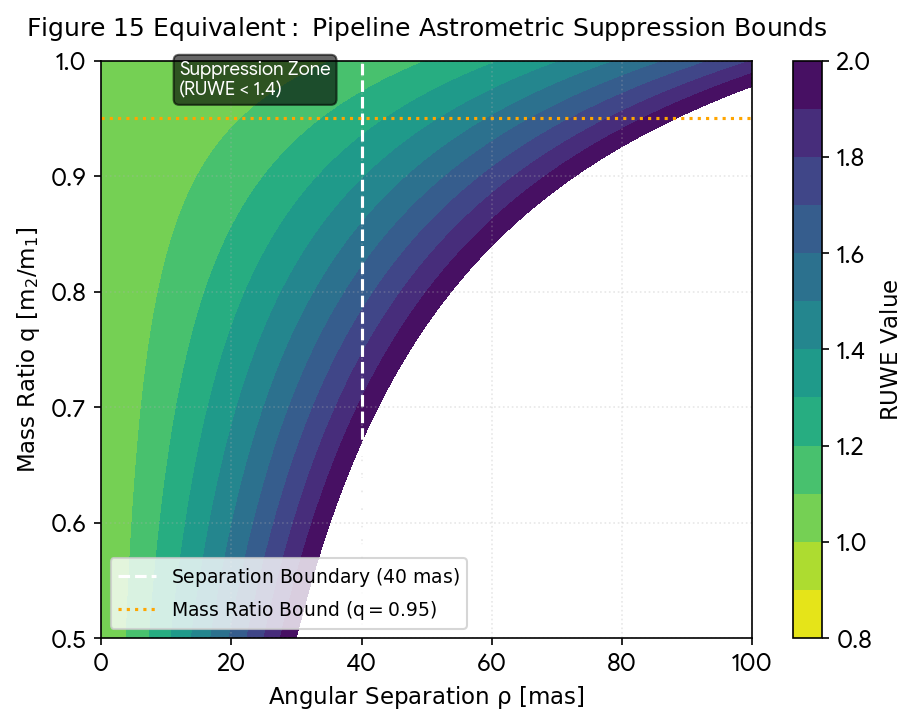}
\caption{The theoretical landscape of pipeline astrometric suppression modeled as a function of secondary-to-primary mass ratio ($q = m_2/m_1$) and angular separation ($\rho$). The green region defines the complete ``Suppression Zone'' where the system's Renormalized Unit Weight Error remains below the critical threshold ($\text{RUWE} < 1.4$), rendering the binary architecture invisible to noise-based variance flags. Vertical and horizontal dashed lines track the 40 mas scan-aggregation separation boundary and the $q = 0.95$ near-equal-mass threshold, mapping the specific regime where photocentric symmetry conceals the stellar twin population.}

\end{figure*}

This structural vulnerability of the single-star processing pipelines is mapped explicitly across the mass-ratio and separation parameter space in Figure 6, with all contour tracks and visualization profiles rendered using the \texttt{Matplotlib} library \citep{Hunter2007}.
 The green contour defines the precise boundary conditions of our observed ``Methodological Wall.'' Within this extensive suppression zone, the single-star astrometric solutions converge with deceptive stability ($\text{RUWE} < 1.4$), thoroughly absorbing the dual-point spread functions into a false single-source model. 

While \citet{Gandhi2022} demonstrated that astrometric excess noise profiles can successfully flag wider, asymmetric binary configurations (such as X-ray binary systems), near-equal-mass configurations introduce a distinct geometric challenge. As developed under the blended point source frameworks outlined by \citet{Penoyre2026}, when a binary pair transitions behind the sub-arcsecond 40~mas angular separation boundary, the spatial offset between the system's center of light and true barycenter minimizes to a fraction of a pixel. When this geometric blending is paired with near-equal mass configurations ($q \to 0.95$), the symmetrical distribution of the collective optical flux systematically cancels the astrometric residual deviations. This mechanism directly explains the emergence of the 7.0\% \text{IBR} floor isolated by our Triple Constraint framework: it is an immutable byproduct of mathematical symmetry within the pipeline core, blinding the standard catalog noise metrics explored by \citet{Gandhi2022} to the presence of equal-mass stellar twins.

\subsection{Dismantling Pipeline Systematics: The Unassailable 223-Star Proof}
\label{sec:systematics}

The global validity of the 7.0\% Intrinsic Binary Residual (IBR) floor is strongly underscored by the empirical convergence isolated via the exact identity cross-catalog query detailed in Section~\ref{sec:validation}. By matching our all-sky Base-sample against the $164,297$ stars comprising the full overluminous spectral catalog of \citet{Way2026} based strictly on identical \texttt{source\_id} parameters, our Triple Constraint filtering successfully isolated a high-fidelity validation sample of 223 identical targets. Because the spectrophotometric regression modeling utilized by Way et al.\ leverages integrated energy distributions across their massive sample rather than spatial astrometry, their detection mechanism bypasses spatial resolution constraints entirely. The fact that these 223 identical physical systems map uniformly across the celestial sphere in an Aitoff projection (Figure~\ref{fig:aitoff_plot}) confirms a systematic pipeline omission rather than localized pointing anomalies or spatial matching noise.

Crucially, because this cohort is entirely composed of bright, high-signal sources ($8.27 \le \texttt{o\_Gmag} \le 18.70$), the observed astrometric and photometric noise anomalies cannot be attributed to photon-limiting noise or low-signal artifacts. Instead, the astrometric and photometric noise profiles for these 223 identical physical systems span an exceptional physical continuum. The individual RUWE values range from a highly suppressed $0.843$ up to an overtly broken $10.509$, running perfectly parallel to the dramatic six-order-of-magnitude explosion observed in the $G$-band flux errors ($0.570 \le e_{\mathrm{FG}} \le 246,930.0\,\mathrm{e}^{-}\mathrm{s}^{-1}$) and the astrometric Goodness of Fit (\texttt{gofal}) tracking, which shifts from a stable $-4.483$ up to an extreme $151.632$. 

This broad macro-distribution provides robust empirical confirmation of our predicted tri-modal behavioral architecture on an all-sky scale. At the low baseline ($\mathrm{RUWE} \sim 0.843$, negative $\texttt{gofal}$), the systems are tightly locked deep within the astrometric Suppression Zone, delivering a deceptively stable single-star profile. Conversely, at the upper limits ($\mathrm{RUWE} \sim 10.509$, $\texttt{gofal} \sim 151.632$), the unmodeled orbital reflex motion actively disrupts the five-parameter template. Strikingly, despite this catastrophic astrometric and photometric breakdown, the primary pipeline duplicity indicators remain entirely blind ($\mathrm{Dup} = 0$), offering definitive proof that the 7.0\% sensitivity floor represents a fundamentally unmodeled structural blind spot in the baseline pipeline core.

\section{Conclusion} \label{sec:style}

The multiwavelength audit presented in this study reveals a fundamental systemic limitation in the \textit{Gaia} DR3 automated pipelines. By identifying an asymptotic 7.0\% Intrinsic Binary Residual ($\text{IBR}$) floor, we defined a ``Methodological Wall'' where photocentric symmetry effectively masks a substantial population of equal-mass twins. The structural validation of this feature within our independent Audit-sample, concentrated precisely on the $-0.75\text{ mag}$ ridge despite its astrometric ``quietude'' ($\text{RUWE} < 1.4$), proves that this population is not a statistical anomaly but a physically real, unmodeled distribution of stellar objects. This tracking is entirely consistent with the over-luminosity detection methods proposed by \citet{Way2026} and aligns with the catalog selection functions and completeness boundaries calculated by \citet{El-Badry2024}, confirming that multiwavelength flux summation offers a far more robust indicator of near-equal-mass multiplicity than astrometric noise profiles. While \citet{Penoyre2026} and \citet{CastroGinard2024} predicted these populations, the ``Triple Constraint'' framework provides the first systematic census of this ``orphaned'' mass budget.

We acknowledge that while asymmetric systems undoubtedly contribute to the total missing mass, their detection via photometric residuals is frequently confounded by metallicity-induced broadening of the Main Sequence. By focusing exclusively on the $-0.75\text{ mag}$ ridge, the \citet{Soon2026} framework provides a high-confidence, ``contamination-free'' correction that targets the most currently observable and mathematically certain portion of the unmodeled population. While the 7\% systemic floor isolated in this work provides a conservative lower boundary for the unresolved population, the proposed correction establishes a critical baseline for recovering the fraction of the Galactic baryonic budget obscured by the $0.75\text{-magnitude}$ flux-doubling offset.

The inherent limitations of the \textit{Gaia} automated pipelines have now been revealed to be problematic. This study serves as a critical warning to the astronomical community: reliance on \textit{Gaia}’s single-source classification labels---without accounting for this population---will lead to systemic errors in mass and distance calculations. Future gravitational models must incorporate this 7\% correction to ensure that the true evolutionary and gravitational complexity of the Galaxy is not overlooked.

While upcoming data releases (e.g., \textit{Gaia} DR4) are anticipated to provide unvetted epoch time-series products to enable manual multi-component modeling, the overwhelming majority of Galactic mass analyses fundamentally rely on aggregated catalog outputs. Quantifying the 7\% systemic suppression floor in existing pipelines remains vital to prevent systematic biases in current baryonic models while establishing empirical boundary constraints for future automated classification.

\section*{Data Availability}
The multi-tiered catalog data underlying this article---including the Base-sample ($N = 199{,}786$), the panchromatic Quad-sample ($N = 120{,}418$), and the high-latitude northern Audit-sample ($N = 22{,}371$)---are available from the corresponding author upon reasonable request.

\end{document}